# Enhanced Intervalley Scattering of Twisted Bilayer Graphene by Periodic AB Stacked Atoms

Lan Meng[1], Zhao-Dong Chu[1], Yanfeng Zhang[2], Ji-Yong Yang[1], Rui-Fen Dou[1], Jia-Cai Nie[1], and Lin He[1],*

[1] Department of Physics, Beijing Normal University, Beijing, 100875, People's Republic of China

[2] College of Engineering, Peking University, Beijing, 100871, People's Republic of China

The electronic properties of twisted bilayer graphene on SiC substrate were studied via combination of transport measurements and scanning tunneling microscopy. We report the observation of enhanced intervalley scattering from one Dirac cone to the other, which contributes to weak localization, of the twisted bilayer graphene by increasing the interlayer coupling strength. Our experiment and analysis demonstrate that the enhanced intervalley scattering is closely related to the periodic AB stacked atoms (the A atom of layer 1 and the B atom of layer 2 that have the same horizontal positions) that break the sublattice degeneracy of graphene locally. We further show that these periodic AB stacked atoms affect intervalley but not intravalley scattering. The result reported here provides an effective way to atomically manipulate the intervalley scattering of graphene.

## I. INTRODUCTION

Owning to its two-dimensional (2D) honeycomb lattice with two equivalent lattice sites (denoted A and B), graphene has two inequivalent Dirac cones at opposite corners of the Brillouin zone, commonly called $K$ and $K'$, see Fig. 1(a). This produces a pseudospin degree of freedom (called as valley isospin) in addition to the electron spin and gives rise to the chirality in the graphene carrier dynamics.[1-7] The presence of the valley isospin and chirality of graphene is of great importance in its transport properties and dominates quantum interference in graphene.[8-15] For example, weak antilocalization resulting from a destructive interference of quasiparticles was expected to be observed in graphene samples without intervalley scattering and chirality breaking scattering.[8,13] Generally speaking, intervalley scattering from one Dirac cone to the other requires a large momentum transfer, which indicates that the valley isospin is, to some extent, robust against disorder in graphene. This offers the possibility to use valley isospin to develop concept devices in the so-called valleytronics.[16,17] Experimentally, intervalley scattering is usually induced by atomically sharp defects, such as the edges of the sample. This elastic scattering allows counterpropagating electrons in graphene to occupy different valleys and results in constructive interference of electrons, which is detected by a suppression of the weak antilocalization and simultaneously a restoration of weak localization.[11-15]

For graphene bilayer with the most common (AB or Bernal) stacking, A and B′ atoms have the same horizontal positions (here the two sublattices in layer 1 and 2 are denoted by A, B and A′, B′, respectively). Then the sublattice degeneracy is lifted and only weak localization is expected and observed.[9,18,19] Assuming a rotation through an angle θ about a B site (directly opposite an A′ atom in the horizontal direction), a commensurate structure is obtained and periodic Moiré patterns can be observed if a B atom is moved to a position of an A′ atom in the horizontal direction, as shown in Fig. 1(b). With a finite interlayer electron hopping, the sublattice degeneracy of the twisted bilayer graphene is partially lifted at these positions that the A(or B) atoms and B′(or A′) are directly opposite in the horizontal direction, i.e., the intervalley scattering is expected to be much enhanced around these atoms. In this paper, we report the enhanced intervalley scattering, which is manifested by weak localization, of twisted bilayer graphene by increasing the interlayer coupling strength. By combined use of transport measurements and scanning tunnelling microscopy (STM), we show that the enhanced intervalley scattering mainly arises from the periodic AB stacked atoms that break the sublattice degeneracy locally. Our experimental result further reveals that these periodic AB stacked atoms do not affect intravalley scattering of graphene. This opens intriguing opportunities to atomically manipulate the intervalley scattering of graphene.

## II. EXPERIMENT

Epitaxial graphene bilayer was grown in ultrahigh vacuum by thermal Si sublimation on hydrogen etched insulating 6H-SiC(000-1), which was reported in a previous paper.[20] According to STM measurements, the epitaxial graphene was mainly bilayer with a twist angle ~ 4.5°.[20] The sample was also characterized by Roman spectroscopy (not shown). The G band and 2D band are located at 1605 and 2686 cm$^{-1}$, respectively, which consist well with that of the bilayer graphene grown on C-SiC substrate. In this work, the same epitaxial graphene bilayer was characterized firstly by STM, scanning tunneling spectroscopy (STS), and transport measurements. Then single-molecule magnets (SMMs) [Mn$_{12}$O$_{12}$(CH$_3$COO)$_{16}$(H$_2$O)$_4$]·2CH$_3$COOH·H$_2$O (Mn$_{12}$-ac), which were synthesized according to Ref. 21, were chemisorbed onto the surface of the epitaxial graphene and

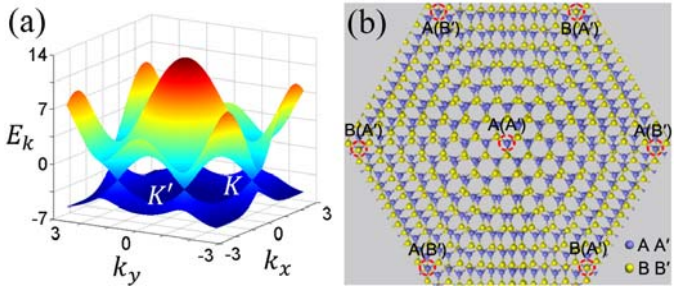

**FIG. 1** (color online). (a) Electronic band structure of graphene. The conduction band and the valence band touch each other at six discrete points, which can be devided into two inequivalent sets, commonly called $K$ and $K'$ (the valley isospin degree of freedom in graphene). (b) Structural model of two misoriented honeycomb lattices with a twist angle that satisfies a condition for commensurate periodic structure leading to Moiré patterns. The two sublattices in layer 1 and 2 are denoted by A, B and A′, B′, respectively.

the obtained sample was further studied by STM, STS, and transport measurements. By adsorption of $Mn_{12}$-ac on the twisted bilayer graphene, we demonstrated that the magnitude of the interlayer coupling is enhanced. The adsorption of $Mn_{12}$-ac may induce local curvature variation of graphene, which is expected to shorten the distance between bilayer graphene and enhance the interlayer electron hopping locally. The low-energy Van Hove singularities (VHSs),[22-24] induced by the interlayer coupling, are observed as pronounced peaks in the tunneling spectra. The interlayer hopping strength is reflected by the separation of the two VHSs flanking the Dirac point. In our experiment, all the STM and STS measurements were performed in an ultrahigh vacuum chamber ($10^{-10}$ Torr) of four-probe SPM from UNISOKU. The STM images were taken in a constant-current scanning mode at liquid-nitrogen temperature (78 K). The STM tips used were made by chemical etching from a wire of Pt(80%)Ir(20%) alloys. The tunneling conductance, *i.e.*, the dI/dV-V curve, was carried out with a standard lock-in technique using a 987 Hz a.c. modulation of the bias voltage. The magnetotransport measurements were carried out using the standard four-point method. The width of the sample is about 3.0 mm and the two inner electrodes were separated about 1.5 mm. A magnetic field was applied perpendicular to the surface of graphene. All the measurements were repeated at different temperatures between 8 and 58 K. Because of the large sample size, our measurements are not obscured by mesoscopic effects and we do not observe any conductance fluctuations.[11-13]

## III. RESULTS AND DISCUSSION

Figure 2(a) shows a STM topography of the epitaxial graphene bilayer grown on SiC. Clear periodic protuberances observed in Fig. 2(a) are Moiré patterns, which arise from misorientation between the top graphene layer and underlying graphene layer.[20] The period of the Moiré pattern is about $D \sim 3.1$ nm. The Moiré pattern can be found in many different regions of the epitaxial graphene, indicating that the main epitaxial graphene is

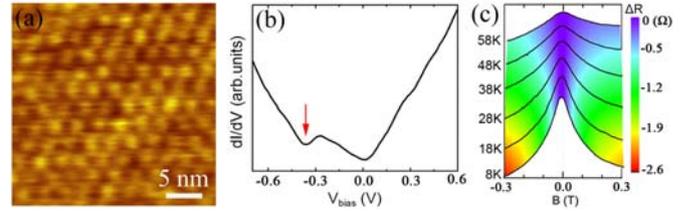

**FIG. 2** (color online). (a) STM image of the epitaxial graphene on SiC ($V_{sample}$ = -480 mV and I = 0.37 nA). The period of the Moiré patterns is about 3.1 nm. (b) A typical dI/dV-V curve obtained on the surface of the epitaxial graphene. The red arrow points to a local minimum (the Dirac point of graphene) $\sim$ -0.40 V in the tunneling conductance spectrum. (c) Weak localization peaks observed in the twisted bilayer graphene at different temperatures (the Y-axis). The color bar in the right represents the variation of magnetoresistance.

bilayer with a twist angle $\theta \sim 4.5°$. Figure 2(b) shows a typical dI/dV-V curves on the epitaxial bilayer graphene. The tunnelling spectrum gives direct access to the local density of states (LDOS) of the surface at the position of the STM tip. Because of charge transfer from the SiC substrate, the epitaxial graphene is intrinsically electron doped. A local minimum (as labeled by the red arrow) at about -0.4 V of the tunneling spectrum is the Dirac point $E_D$ of the graphene. The electron doping level is similar to that reported in literature.[25,26] Here we should pointed out that the absence of VHSs reported here is not related to the experimental temperature, $\sim$ 78 K. Similar experiments done on twisted graphene at 10 mK also did not see any VHSs.[27] The VHSs appear only when there is finite interlayer coupling between twisted graphene bilayer.[22-24]

Figure 2(c) shows the magnetoresistance (MR), $\Delta R = R(B) - R(0)$, of the twisted bilayer graphene measured at different temperatures. Negative MR corresponding to weak localization is observed. The slight asymmetry of the MR about zero magnetic field may arise from asymmetry of electrodes. The MR behavior of graphene, which originates from quantum interference of quasiparticles, can be described by[10-15]

$$\Delta R(B,T) = -\frac{e^2 R^2}{\pi h}\left[ F\left(\frac{B}{B_\phi}\right) - F\left(\frac{B}{B_\phi + 2B_i}\right) - 2F\left(\frac{B}{B_\phi + B_*}\right)\right]. \quad (1)$$

Here $F(z) = \ln(z) + \psi(0.5+z^{-1})$, $\psi(x)$ is the digamma function, $\tau_{\phi,i,*} = h/(8\pi De B_{\phi,i,*})$, and the diffusion coefficient $D = v_f l/2$ with $l$ the mean free path.[14] $\tau_\phi$ represents dephasing time of inelastic scattering, $\tau_i$ and $\tau_*$ depict elastic process of intervalley scattering and intravalley scattering respectively. The Eq. (1) was originally developed for monolayer graphene[10] and subsequently extended to the Bernal stacked bilayer graphene[19] with interlayer coupling of 0.27 eV.[23] Therefore, it is reasonable to expect that the Eq. (1) is applicable to describe the magnetoresistance of graphene bilayer without interlayer coupling, as shown in Fig. 2, and of graphene bilayer with a finite interlayer coupling of 0.07 eV, as shown in Fig. 3. For strong intervalley and

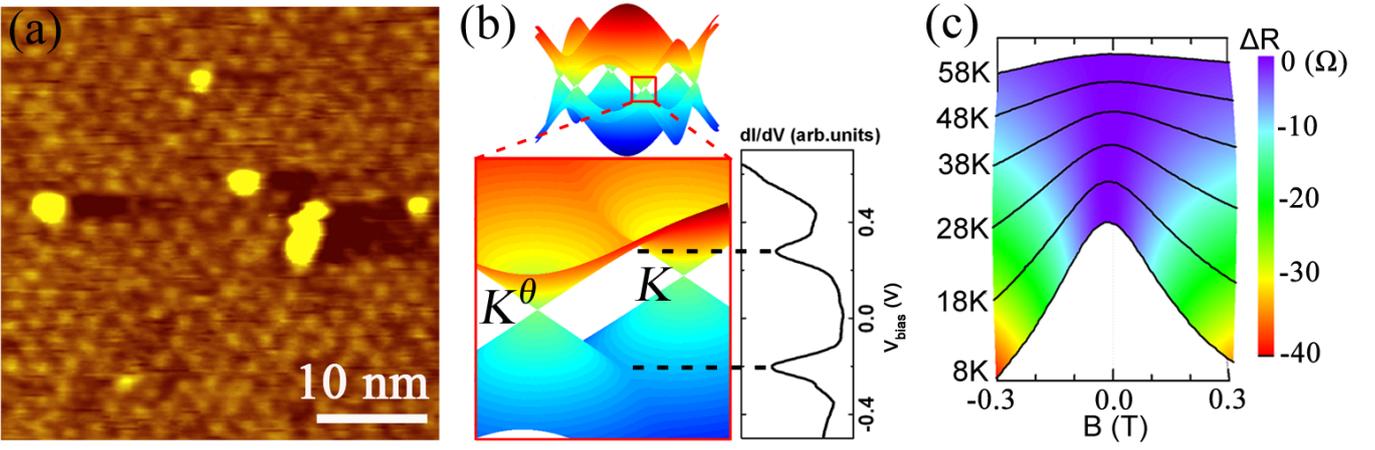

**FIG. 3** (color online). (a) A typical STM image of $Mn_{12}$-ac molecules adsorption on the surface of the twisted bilayer graphene ($V_{sample}$ = 1.2 V and I = 0.10 nA). The bright yellow dots are $Mn_{12}$-ac molecules. (b) Electronic band structure of twisted bilayer graphene with a finite interlayer coupling calculated with the four-band model. Two saddle points (VHSs) form at $k = 0$ between the two Dirac cones, $K$ and $K_\theta$, with a separation of $\Delta K = 2K\sin(\theta/2)$. The low-energy VHSs contributes to two pronounced peaks flanking zero-bias in a typical tunneling spectrum obtained in our experiment. (c) Weak localization peak at different temperatures (the Y-axis) observed in the twisted bilayer graphene with adsorption of $Mn_{12}$-ac molecules. The color bar in the right represents the variation of magnetoresistance.

intravalley scattering, the first term, which is responsible for the weak localization, dominates. In the opposite case of negligible intervalley and intravalley scattering, the MR is totally controlled by the terms with negative sign, which result in weak antilocalization. Figure 4 summarizes the fitting parameters to the experimental data according to Eq. (1). Three parameters $\tau_\phi$, $\tau_i$, and $\tau_*$ are adjustable in the calculation. Further discussion about $\tau_\phi$, $\tau_i$, and $\tau_*$ will be elaborated subsequently.

The adsorption of $Mn_{12}$-ac on the surface of the twisted bilayer graphene may result in local curvature variation of graphene and therefore tune its interlayer hopping and electronic band structure. Figure 3(a) shows a typical STM topography of the sample. The $Mn_{12}$-ac molecules with the size of ~ 2.0 nm disperse randomly on the surface of graphene. Due to the enhanced interlayer hopping, two VHSs form at $k = 0$ between the two Dirac cones, $K$ and $K_\theta$, with a separation of $\Delta K = 2K\sin(\theta/2)$,[22-24] which contribute to two pronounced peaks flanking zero-bias in the tunneling spectrum, as shown in Fig. 3(b). To confirm the observed phenomena, we carried out the STS measurements at different positions of the graphene. The main features of the STS curves are completely reproducible. The electronic band structure of the twisted bilayer graphene was calculated by tight binding model with the Hamiltonian

$$H = H_1 + H_2 + H_\perp, \quad (2)$$

where $H_1 = -t\Sigma_{<i,j>}(c_{Ai}^+ c_{Bj} + H.c.)$ and $H_2 = -t\Sigma_{<i,j>}(c_{A'i}^+ c_{B'j} + H.c.)$ are the Hamiltonians for each layer, and $H_\perp = \Sigma_{\alpha,\beta}^i t_\perp^{\alpha\beta}(r_i) c_\alpha^+(r_i) c_\beta(r_i + \Delta r_i)$ is the interaction Hamiltonian between the two layers.[22,25] $\Delta r_i$ is the horizontal (in-plane) displacement from an atom of layer 1 to the closest atom in layer 2, the sublattice $\alpha$ = A, B, and $\beta$ = A′, B′, $t_\perp^{\alpha\beta}(r_i)$ is the hopping amplitude between nearest neighbor atoms from different layers. To describe the electronic band structure of twisted graphene bilayer in the whole energy interval rather than limited in the low energy around the Dirac points, we assume that $t_\perp^{\alpha\beta}(r_i)$ is a constant, $t_\perp^{\alpha\beta}(r_i) \approx t_\perp$ in the calculation. Using the standard replacement, $c(r_i) = \Sigma_k(e^{-ik\cdot r_i})\psi_k$. The energy spectrum can be obtained by diagonlizing the matrix of Hamiltonian $H$, as shown in Fig. 3(b). The twist angle $\theta$ ~ 4.5° and the interlayer hopping parameter $t_\perp$ ~ 70 meV are used in the calculation of the band structure (The interlayer hopping parameter of the sample can be estimated by $\Delta E_{vhs} = \hbar v_F \Delta K - 2t_\perp$.[23] Here, $\Delta E_{vhs}$ is the energy difference of the two VHSs, $v_F$ is the renormalized Fermi velocity of bilayer graphene with a twist angle). The obtained low-energy electronic band structure consists with that of the effective two-band (four-band) model in the continuum approximation [22,28] and agrees quite well with the result obtained from STS measurement. This indicates that the interlayer hopping parameter of the graphene bilayer increases from almost zero [29] to about 70 meV with absorption of $Mn_{12}$-ac.

According to the analysis at the beginning, the interlayer coupling will lift the sublattice degeneracy of the twisted bilayer graphene at the periodic AB stacked atoms. Therefore, the intervalley scattering and consequently the weak localization of the sample is expected to be enhanced. Figure 3(c) shows the MR of the sample measured at different temperatures. Obviously, the weak localization, i.e., negative MR, is much enhanced if comparing with that shown in Fig. 2(c). The fitting parameters to the experimental data according to Eq. (1) are also plotted in Fig. 4. The intravalley scattering time $\tau_*$ ~ 0.03 ps, which is independent of temperature, of graphene is almost not affected by the absorption of $Mn_{12}$-ac molecules. However, the intervalley scattering time $\tau_i$ of graphene is twice as that of graphene after the absorption. Consequently, the intervalley scattering rate $\tau_i^{-1}$ of the twisted bilayer graphene with a finite interlayer hopping is twice of that without interlayer coupling. This result indicats that the newly added "defects" after

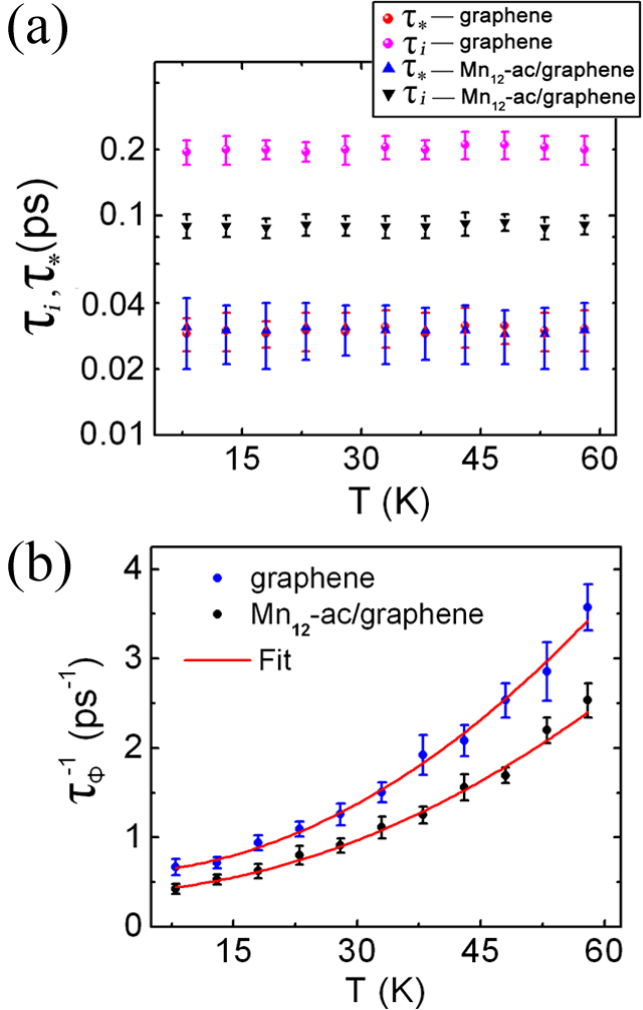

**FIG. 4** (color online). (a) Temperature dependence of the intravalley scattering time and the intervalley scattering time before and after Mn$_{12}$-ac molecules absorbed onto the twisted graphene bilayer. (b) Temperature dependence of the dephasing rate of the two samples. The solid curves are fitting results of Eq. (3). The solid symbols are the best fit values for the different scattering times and the error bars represent the possible minimum and maximum parameters that can give a well fit to our experimental data.

adsorption of Mn$_{12}$-ac affect intervalley but not intravalley scattering. This is easy to understand since these "defects" originate from sublattice degeneracy breaking at these positions that the A(or B) atoms and B′(or A′) are directly opposite in the horizontal direction. Our experimental result also reveals that the adsorbed Mn$_{12}$-ac molecules did not direct scatter electrons in graphene.

Figure 4(b) summarizes the temperature dependence of the dephasing rate of the twisted bilayer graphene both before and after the adsorption of Mn$_{12}$-ac. The slight difference between the dephasing rates may partly arise from local curvature variation of graphene induced by the adsorption. The local curvature variation is expected to shorten the distance between bilayer graphene locally to enhance the interlayer electron hopping. Usually, the dephasing scattering in graphene is dominated by the electron-electron interaction.[13,15] The temperature dependence of the dephasing rate in graphene should be expressed as[30,31]

$$\tau_\phi^{-1} \propto \tau_N^{-1} + \tau_{ee}^{-1} + c. \quad (3)$$

Here $\tau_N^{-1} = a k_B T \frac{\ln(g)}{\hbar g}$ corresponds to the inelastic scattering with a small momentum transfer, which mainly originates from interaction of an electron with the fluctuating electromagnetic field generated by noisy movement of neighboring electrons, $\tau_{ee}^{-1} = b \frac{\sqrt{\pi}}{2 v_F} \left( \frac{k_B T}{\hbar} \right)^2 \frac{\ln(g)}{\sqrt{n}}$ depicts a large-momentum-transfer induced by direct Coulomb interaction, $g(n) = \sigma(n) h / e^2$ is the normalized conductivity, and $c$ is a constant. The fitting results of the experimental data to Eq. (3) are shown in Fig. 4. The parameters $a$, $b$, and $c$ are adjustable in the calculation. Our analysis indicates that the contribution to the dephasing scattering rate in graphene mainly arises from electron-electron interaction, which agrees quite well with that reported in literature.[14,15] Additionally, similar temperature dependence of the dephasing rate in graphene, as depicted by Eq. (3), was attributed to the presence of local magnetic moments by introducing fluorine adatoms into graphene.[32] In our samples, the local magnetic moments may arise from atomic defects in graphene.[33]

To further explore the effect of the adsorption of Mn$_{12}$-ac, we also deposited Au nanoparticles with diameter of about 2.8 nm on the graphene grown on SiC substrate.[34] The deposited Au nanoparticles almost not affect the electronic properties of the graphene. In this work, we found that the Dirac point of the as-grown graphene is lifted from -0.4 eV (Fig. 2(b)) to about 0 eV (Fig. 3(b)) by adsorption of Mn$_{12}$-ac. It indicates that there is charge transfer between Mn$_{12}$-ac and graphene, which may be the possible origin of that the adsorption of Mn$_{12}$-ac tunes the interlayer hopping and electronic band structure of twisted graphene bilayer. However, the exact origin is still unknown at present.

## IV. CONCLUSIONS

In summary, we studied the electronic properties and structures of twisted bilayer graphene on SiC substrate via transport measurements and STM. By enhancing interlayer electron hopping, we demonstrate that the intervalley scattering and the weak localization could be enhanced. The enhanced interlayer scattering is attributed to the breaking of sublattice degenerancy at the positions that the A(or B) atoms of top layer and B′(or A′) of the sublayer are directly opposite in the horizontal direction. Recently, it is predicted that the interlayer coupling of twisted bilyer graphene with a small twist angle could give rise to flat bands near Fermi level.[35-37] A combined use of STM and transport measurements to study such system could be helpful to explore many attractive properties, such as novel superconductivity,[38,39] in graphene bilayer in the near future.

## ACKNOWLEDGEMENTS

This work was supported by the National Natural Science Foundation of China (Grant Nos. 10804010, 10974019, 11004010 and 21073003), the Fundamental Research Funds for the Central Universities, and the Ministry of Science and Technology of China (Grants Nos. 2011CB921903).

*Email: helin@bnu.edu.cn

666